\documentclass[12pt]{article}
\usepackage{amsmath,amssymb,graphicx,mathrsfs,hyperref,slashed,setspace}

\usepackage{float}
\usepackage{latexsym}
\usepackage{bm}
\usepackage{blindtext}
\hypersetup{
	colorlinks=true,
	linkcolor=blue,
	filecolor=black,
	urlcolor=black,
	citecolor=blue,
}

\newcommand{\hide}[1]{}

\renewcommand{\baselinestretch}{1.5}

\newcommand{\be}{\begin{equation}}
\newcommand{\ee}{\end{equation}}
\newcommand{\bea}{\begin{eqnarray}}
\newcommand{\eea}{\end{eqnarray}}

\def\({\left(} \def\){\right)}

\renewcommand{\baselinestretch}{1.5}
\begin{document}
\title{\vspace{-1.8in}
{Sourcing the Kerr geometry}}
\author{\large Ram Brustein${}^{(1)}$,  A.J.M. Medved${}^{(2,3)}$
\\
\vspace{-.5in} \hspace{-1.5in} \vbox{
\begin{flushleft}
 $^{\textrm{\normalsize
(1)\ Department of Physics, Ben-Gurion University,
   Beer-Sheva 84105, Israel}}$
$^{\textrm{\normalsize (2)\ Department of Physics \& Electronics, Rhodes University,
 Grahamstown 6140, South Africa}}$
$^{\textrm{\normalsize (3)\ National Institute for Theoretical Physics (NITheP), Western Cape 7602,
South Africa}}$
\\ \small \hspace{0.57in}
   ramyb@bgu.ac.il,\  j.medved@ru.ac.za
\end{flushleft}
}}
\date{}
\maketitle

\renewcommand{\baselinestretch}{1.15}
\begin{abstract}

The Kerr metric is a vacuum solution of the Einstein equations outside of a rotating black hole, but what interior matter is actually rotating and sourcing the Kerr geometry?  Here, we describe a rotating exotic matter which can source the Kerr geometry for the entire acceptable range of its spin parameter and be shown to saturate the radial null-energy condition at every point in the interior, while being free of any obvious pathologies. We do so by introducing the rotating  frozen star, whose compactness is controlled by a perturbative parameter and whose outer surface can be arbitrarily close to the horizon of a Kerr black hole. The interior geometry modifies Kerr's such that there is neither an inner ergosphere nor an inner horizon, and the metric and Einstein tensors are regular everywhere except for a mild, removable singularity at the center of the star.  The geometry of each radial slice of the interior is a nearly null surface with the same geometry, but different radial size, as  that of the  would-be horizon on the outermost slice. Moreover, the metric limits to that of the static frozen star as the spin is taken to zero.  The integral of the energy density leads to a rest mass that is equal to the  irreducible mass of a Kerr black hole, and the integral of the angular-momentum density confirms that the ratio of the angular momentum to the mass is equal to the Kerr spin parameter. Including the rotational energy in the standard way, we then obtain the total gravitational  mass and angular momentum of a Kerr black hole with the same mass and spin parameters. We propose that the rotating frozen star provides a classical description of the highly quantum final state of gravitationally collapsed matter, which features a maximally entropic, stringy fluid as described by our polymer model for the black hole interior.

\end{abstract}
\maketitle

\newpage
\renewcommand{\baselinestretch}{1.5}
\section{Introduction}

Most astrophysical black holes (BHs) are rapidly rotating objects, whose exterior geometry is believed to be described by the Kerr solution. But what is the interior matter that is actually rotating and sourcing the Kerr geometry?

This question has attracted much attention over the years, starting at the time that the Kerr solution was found \cite{KerrLecture} up to the present day. However, many of the numerous attempts at modeling a regularized mimicker as an alternative to a general-relativistic BH  discuss the static case (see \cite{carded} for a comprehensive catalogue). Attempts at including rotation are, in many instances, limited to slowly rotating versions of  already-existent static models or put some limitations on the possibility of finding  rotating mimickers that can maintain regularity. Other models allow some pathologies, such as violations of  the averaged null-energy condition or strong instabilities, with the hope that such ailments will be cured by quantum effects. A review of the earliest  attempts can be found in \cite{W0} and a more up-to-date review can be found in \cite{Y0}. A partial list of relevant discussions can be found in \cite{X0,X00,X000,X1,X2,X3,X4,X5,X6,X7,X8,X9,X10,X11,X12,X13,X14,X15,X16,X17,X18,Y1,Y2}.

Here, we propose that a possible answer to this question is  a rotating frozen star. This is an ``upgrade''  of our proposed model for a BH mimicker \cite{bookdill,BHfollies},  recently dubbed as
the frozen star \cite{popstar} (also see \cite{trajectory,fluctuations,U4Euclidean}), although
there is no claim of originality in the use of this name \cite{RW}.  This model is meant to represent what the interior, classical geometry of a regular, non-rotating BH might look like.   Moreover, it is  meant  to provide an effective {\em geometric description} of the equilibrium state of a  collection of highly excited, long, closed, fundamental strings. Meanwhile, our  original proposal for describing  the {\em quantum state} of such strings \cite{inny} --- the polymer model --- has been shown to
exhibit   all of the classical \cite{strungout} and quantum \cite{emerge} properties of an unperturbed  Schwarzschild BH.  What would distinguish the polymer from an actual  BH is  its out-of-equilibrium behavior, as this cannot be explained without some knowledge of  the object's  fundamental
composition \cite{ridethewave,fluctuations}.

The underlying polymer model is based  on a sequence of well-motivated assumptions \cite{inny}: (1)~To avoid contradictions and paradoxes, a BH must deviate significantly from its general-relativistic description. This can either be in the form of a so-called firewall that is localized at its surface \cite{AMPS} or on horizon-sized length scales within. (2)~The latter option requires a strongly non-classical state of matter throughout the interior, which could evolve from normal matter if it undergoes a quantum-gravitational phase transition or  tunneling event, as described in \cite{mathur,U4Euclidean}. (3)~Extreme non-classicality is synonymous with maximal entropy \cite{CEB}. (4)~A fluid with maximal entropy density can only be realized  by  a theory of  highly excited,  long, closed  strings, as this matter alone is on the cusp of violating causality,
$\;p = \rho\;$ \cite{AW}, meaning that the entropy density  $s$ at fixed temperature $T$ and fixed energy density $\rho$ cannot be made any higher, $\;s=\frac{p+\rho}{T}\;.$
Here and onward, $p$  refers to the radial pressure as a theory of long, closed strings is effectively $1+1$-dimensional.

Back to our classical solution, maximally negative radial pressure  $\;p=-\rho\;$, was found to be a key ingredient in attaining the condition of regularity at a BH horizon  \cite{MMV1,MMV2} and then, by extension, throughout the interior of the frozen star. This condition corresponds to the saturation of the radial component of the null-energy condition (RNEC), which is central to the singularity  theorems \cite{PenHawk1,PenHawk2} and essential for circumventing the closely related  bounds that limit the  compactness of gravitationally collapsing systems \cite{Buchdahl,chand1,chand2,bondi,MM}.

The link between the quantum and classical pictures now becomes clear. Maximally positive pressure and, thus,  maximal entropy tells us how far causality can be pushed  before spacetime ceases to have a semiclassical description. Maximally negative pressure and, thus, the saturation of the  RNEC also tells us how far causality can be pushed  but from  a different perspective. The former viewpoint considers the actual matter content while the latter is stated in geometric terms.  Thanks to the saturation of their respective bounds, the two models share another important feature: Each radial slice of the interior is itself  a nearly null surface, just like the  would-be horizon on the outermost slice.

Our own rotating solution of the Einstein equations will be required to have the following properties in addition to regularity:
(1)~The solution should, to leading order in a perturbative parameter,  saturate the RNEC throughout the interior for a class of probe observers that can formally reach the inside of the star and (2)~it should be able to reproduce the total mass and angular momentum of a Kerr BH.  The former requirement ensures its connection to the polymer model, while the latter guarantees that the interior ``fluid'' sources the exterior Kerr  geometry. We will, for the sake of consistency,  also assume that the solution agrees with the static frozen star in the zero-rotation limit.

The paper is organized as follows:  The rotating frozen star is introduced in Section~2.  To this end, we start from the Kerr metric in an ADM-like, near-horizon version of  Boyer--Lindquist coordinates. This form of the metric is then transformed into another that is appropriate  for  a zero-angular-momentum observer (ZAMO). We also make use of a closely related coordinate system  from  \cite{MMV2} for which the  RNEC is exactly saturated on the Kerr horizon. In our case, however,  the same condition is saturated, to leading order in the perturbative parameter,  throughout the interior of the rotating star. This can be seen after finally imposing the transformation from the ZAMO--Kerr geometry to that of the rotating frozen star. The existence of a Killing tensor for the rotating star interior is established in the second major part of Section~2. The Einstein tensor for the just-discussed metric is  calculated  in Section~3; leading in particular  to the energy density and radial pressure from the perspective of  a locally corotating observer.  We integrate this energy  density over the volume of the star and show that the star's rest mass  is equal to the irreducible mass of the corresponding  Kerr BH. We then integrate the angular-momentum density and find that the ratio of the angular momentum to the rest mass is exactly equal to the Kerr spin parameter. The rotational energy is included in the standard way to obtain the total gravitational  mass, which matches
with that of its Kerr BH counterpart. The Kerr value for the angular momentum immediately follows. We further show that the energy density is formally negative in a certain region of the interior. However, from the perspective of the class of probe observers that can access this region, the energy density actually vanishes. The energy density is also mildly singular near the center of the star; however, we expect that it can be regularized by a  procedure which is similar  to that used to regularize non-rotating frozen stars \cite{trajectory}. The paper then ends with a brief overview.

\section{A rotating frozen star: The geometry}

\subsection{The metric}

Our starting point is the Boyer--Lindquist metric for the Kerr spacetime; see, for example, \cite{MTW}.
Near the horizon,  this metric can   be reexpressed  in an  ADM-like form  \cite{MMV2} (see the Appendix for details),
\be
ds^2\;=\; -\left(N^2-\omega^2g_{\phi\phi} \right)dt^2+ \frac{dr^2}{N^2} -2\omega g_{\phi\phi}dtd\phi
+g_{\theta\theta}d\theta^2
 +g_{\phi\phi}d\phi^2
 \;.
\label{metricX}
\ee
Here, $\omega=-\frac{g_{\phi t}}{g_{\phi\phi}}$ can be interpreted as a local measure of the angular velocity and $N$ is a standard lapse function in that $\;N^2=g^{rr}=0\;$ at the horizon or surface of ``infinite redshift''. Let us recall a few properties of the Kerr geometry. The horizon at $\;r=r_+\;$ is defined by $\;\Delta=r^2 -2Mr + a^2=0\;$
 and is located at $\;r_+=m+\sqrt{M^2-a^2}\;$, where $M$ is the ADM mass of the BH and $a$ is the dimensional spin parameter, $\;0\leq a \leq M\;$. The lapse function is $N^2=\Delta/\Sigma$, with $\Sigma= g_{\theta\theta}=r^2+a^2 \cos^2\theta$.  Also,  $\;N^2 =\omega^2 g_{\phi\phi}\;$ at the ergosphere which is situated at
$\;r=r_{ES}=M+\sqrt{M^2-a^2\cos^2{\theta}}\;$. It is worth recalling that $\;r_+<r_{ES}<r_S=2M\;$.
The inner horizon and inner ergosphere are of no interest here,  as these are vanquished for  the rotating frozen star.

Next, we  transform to ZAMO coordinates, which  are those for  a special class of observers for which the locally measured angular momentum vanishes \cite{MTW,frolovtimes2}. This amounts to  adapting the coordinates with respect to the Killing vector
$
\;\chi^a = \xi^a+ \omega\psi^a\;=\;(1,0,0,\omega)\;,
$
where $\xi^a$ and $\psi^a$ are the usual timelike and axially symmetric Killing vectors for the Kerr spacetime.

In terms of the adapted time coordinate,
the line element in the near-horizon region becomes,
\be
ds^2 \;\to\; -N^2 d\chi^2 + \frac{dr^2}{N^2}
+g_{\theta\theta}d\theta^2
 +g_{\phi\phi}d\phi^2
 \;.
\label{diagonal}
\ee
This is precisely the horizon limit of
the metric obtained in \cite{MMV2},
\be
ds^2 \;=\; -\left[N^2 -\left(\Omega_{H}-\omega\right)^2g_{\phi\phi}\right]d\chi^2 + \frac{dr^2}{N^2}
+ \left(\Omega_{H}-\omega\right)g_{\phi\phi}d\chi d\phi     +g_{\theta\theta}d\theta^2
 +g_{\phi\phi}d\phi^2
\;,
\ee
with $\Omega_{H}= \frac{a}{r_+^2+a^2} = \frac{a}{2Mr_+}$ being the horizon value of $\omega$  --- the only angular velocity with which observers on the horizon are allowed to move.

The above metric was used in \cite{MMV2} to show that the RNEC
is exactly saturated on the horizon of a rotating BH.
This happens because, as the horizon is approached,
$\;\Omega_H-\omega\;$ has to go to zero at least as fast as the square of the lapse function, as this ensures the regularity of scalar invariants involving the Riemann curvature.
So that, in this limit, one obtains the diagonal metric~(\ref{diagonal})
for which  the elements of the Einstein tensor satisfy   $\;G^{\chi}_{\;\;\chi}=G^{r}_{\;\;r}\;$
by virtue of the temporal--radial sector of the Einstein tensor being directly proportional
to the  metric. In other words, $\;\rho+p=0\;$.

We now turn our attention to the interior, as the spacetime external to the horizon  will be left alone (but see Footnote~\ref{f1}). The near-horizon form of the metric~(\ref{metricX}) is sufficient for our purposes because one of the premises of the  frozen star model is that, up to perturbatively small corrections, every radial slice of the interior is an almost-null surface whose geometry is like that of a BH horizon. Introducing the perturbatively small parameter $\varepsilon^2= \dfrac{\Delta}{\Sigma}$, we thus write,
\be
ds^2 \;=\; -\varepsilon^2 d\chi^2 + \frac{dr^2}{\varepsilon^2}
+g_{\theta\theta}d\theta^2
+g_{\phi\phi}d\phi^2 \;,
\ee
where $\;\varepsilon^2\ll 1\;$.~\footnote{What we now call $\varepsilon^2$ had, until recently, been called
$\varepsilon$. The change was made to emphasize its positivity.}
To get some idea of just how small, a recent paper used empirical data to argue that
$\;\varepsilon^2\lesssim 10^{-22}\;$ \cite{clowncar}.

To complete the interior metric, we set the angular components of the metric equal to their Boyer--Lindquist counterparts but with one caveat:
The component $g_{\phi\phi}$ depends on the ADM mass but, for the static frozen star, the
mass  was redefined by a scaling relation $\;M\to M(r)=\frac{r}{2}(1-\varepsilon^2)\;$.
The natural generalization for the rotating case is
$\;M(r)=\frac{r^2+a^2}{2r}(1-\varepsilon^2)\;$, which can be used to rewrite $g_{\phi\phi}$ in a suitable ``near-horizon'' form.
This leads to
\be
ds^2 \;\to\; -\varepsilon^2 d\chi^2 + \frac{dr^2}{\varepsilon^2}
+\Sigma d\theta^2 +\frac{(r^2+a^2)^2}{\Sigma} \sin^2{\theta} d\phi^2\;.
\label{diagonal2}
\ee
where we kept only the leading term in $g_{\phi\phi}$.
Our subsequent calculations are, consequently, valid only to leading order in $\varepsilon^2$, but we do expect that it is possible to obtain results that are valid to higher order by adding $\varepsilon^2$ corrections to Eq.~(\ref{diagonal2}).

By design and as an explicit calculation confirms, the RNEC is saturated, to leading order in $\varepsilon^2$, throughout the interior of the rotating star given this choice of metric. A consequence of this saturated condition is that every surface of fixed radial coordinate $r$ has, to leading order, the same geometry as the horizon of a Kerr BH, except that the redshift is large but finite rather than infinite.  The area of this surface is $A= 4\pi R^2=4\pi (r^2+a^2)$, with $R$ being the Kerr areal coordinate.

Besides ensuring the saturation of  the RNEC, the metric~(\ref{diagonal2})
limits to that  of the static frozen star as $\;a\to 0\;$ {\em and}, perhaps most importantly, is regular throughout the interior except for a mild, integrable  singularity at the center of the star. But the constraint that uniquely fixes the form of the metric
is that its corresponding Einstein  tensor is  able to reproduce the mass and angular momentum of a Kerr BH,
as will be shown  in Section~3.

Before moving on to discuss symmetries, there are a few pertinent points to be made.

\begin{enumerate}
\renewcommand{\labelenumi}{\alph{enumi})}

\item
Given the form of the interior metric~(\ref{diagonal2}), one might wonder if there is a natural way to limit the value of $a$ like there is for the original  Kerr solution. However, it must be the case that $\;a\left(r\lesssim r_+\right)\leq a\left(r\gtrsim r+\right)\;$; otherwise, the star will be shedding mass until
an equilibrium state is attained. It then follows that $a_{interior} \leq a_{exterior}\leq M\;$.

\item  Matching the interior and exterior solutions will require a narrow transitional layer  (width $\;\lambda\ll r_+$), which allows for the interior metric and its first two derivatives to be continuously deformed  so as to match up with the exterior geometry, just as for the static frozen star \cite{popstar}.~\footnote{In this regard and also including corrections of order $\varepsilon^2$, the interior geometry formally stops  at $\;r_+ +\varepsilon^2-\frac{\lambda}{2}\;$
and the  exterior geometry formally begins  at  $\;r_+ +\varepsilon^2 +\frac{\lambda}{2}\;$ and neither at $r_+$ as implied  earlier.\label{f1}}
We do not expect the process to be significantly different from that used in \cite{popstar} for
the static case. We also do not expect the inclusion of the layer to modify calculations such as the mass in the next section, at least not at leading order.

\item
On the other hand, as can be seen in Section~3, the curvature and all non-vanishing components of the Einstein tensor remain regular throughout the interior, except for a mild singularity near the center of the star, as in the static case \cite{trajectory}. We also expect, similarly
to the static case, that a regularization zone can be introduced near the center of the rotating star that will affect the star's mass and angular momentum  by  negligibly small amounts.

\item
The property that
ensures the ultrastability of the static frozen star,  $\;G^t_{\;\;t}=G^r_{\;\;r}\;$,  remains intact, and so we anticipate that the interior of the rotating frozen star is similarly ultrastable.
\end{enumerate}

\subsection{Killing vectors and tensor}

It is clear that the  timelike and axially symmetric Killing vectors survive for the revised interior, but what about the Kerr Killing tensor? This is, to some extent, a moot issue inasmuch as  any material body will be torn apart as it approaches the (would be) horizon. So that, when discussing the symmetries of the solution, we need  formally to consider probe observers; for example, a photon that propagates through the interior.

Let us further  point out that  any such ``observer'' must be moving on a radial, null trajectory
in the frozen star interior. To understand why, one can consider
a convenient form of a particle's  equation of motion,
\be
k\;=\; -\varepsilon^2 (u^t)^2+ \frac{1}{\varepsilon^2}(u^r)^2+ g_{\theta\theta}(u^{\theta})^2 +g_{\phi\phi}(u^{\phi\phi})^2\;,
\label{motion0}
\ee
where $u^a$ is a component of the $4$-velocity, while $\;k=1\;$ for a  timelike and $\;k=0\;$ for a null trajectory.

As the asymptotic conserved energy per unit mass is $\;{\cal E}= -g_{tt}u^t=\varepsilon^2 u^t\;$, Eq.~(\ref{motion0})   can be recast as
\be
{\cal E}^2-(u^r)^2\;=\;  \varepsilon^2\left[g_{\theta\theta}(u^{\theta})^2 +g_{\phi\phi}(u^{\phi\phi})^2 +k\right]\;,
\label{motion}
\ee
where $\;{\cal E}\gtrsim \Omega_H L\;$ for a ZAMO
such that  $\;L=g_{\phi\phi}u^{\phi}\;$ is the axial angular momentum per unit mass, which is also conserved.

One can now see that  the angular components of velocity, as well as
the distinction between null and   timelike motion, are suppressed by a factor of  $\varepsilon^2 \;$ throughout the frozen star interior,
\be
{\cal E}^2  - (u^r)^2 \;=\; {\cal O}[\varepsilon^2]\;,
\label{almostnull}
\ee
where the correction represents  a perturbatively small deviation from a perfectly null, radial path. Further note that objects with  large angular velocities at infinity would not get through the external potential barrier, so that the right-hand side will not have large-enough numbers to compensate for
the  factor  of $\varepsilon^2$.

With all  this in mind, let us recall the Killing tensor's standard form   \cite{carter},
\be
K^{ab}\;=\;\Sigma\; l^{\left(a\right.}n^{\left.b\right)}+r^2g^{ab}\;,
\label{KillCarter}
\ee
where $l^a$ and $n^a$
are null and mostly radial vectors with normalization
$\;l^an_a=-2\;$. In the ZAMO case,
$\;l^a=\left(\frac{1}{\varepsilon^2},1,0,0\right)\;$,
$\;n^a=\left(1,-\varepsilon^2,0,0\right)\;$  and the Killing tensor does  not quite satisfy the requisite identity,
\be
I^{abc}\;\equiv\;\nabla^{\left(a\right.}K^{\left.bc\right)}\;=\;0\;,
\label{kill}
\ee
in all instances. It does only fail for two particular cases (and permutations thereof): $\;I^{r\phi\phi}= 4\varepsilon^2\frac{ra^2\Sigma}{(r^2+a^2)^3} \;$  and $I^{r r\theta}=\varepsilon^2\frac{a^2\sin{2\theta}}{\Sigma}\;$.
Both violations are of order $\varepsilon^2$, but cannot  immediately be dismissed  on this basis  since their covariant forms $I_{abc}$ are not similarly  suppressed. We  will rather argue that these apparent violations do not change the integrability of the Hamilton--Jacobi equations nor the constancy of the Carter constant, again, to leading order in $\varepsilon^2$.

Let us recall the Carter constant
\be
C=K_{ab}u^a u^b
\label{Karter}
\ee
and that it is one of the  four conserved quantities for motion in a Kerr spacetime \cite{carter}.
It is, essentially, the square of the total angular momentum  of a propagating particle  {\em plus} a contribution that can be attributed to a distortion in its radial propagation when $\;a\neq 0\;$ \cite{CC}.

 In the case of an exactly radial, null trajectory in the frozen star interior, one finds from Eq.~(\ref{KillCarter}) that
\be
C\;=\;K_{ab}u^a u^b\;=\;0\;.
\ee
The Carter constant in this case is not only zero but identically vanishing, and so its derivatives  are also identically vanishing. This result is a consequence of  the structure of the $\chi$--$r$ sector of the geometry, thus rendering $I^{r\phi\phi}$ and $I^{r r \theta}$  as irrelevant to the constancy of the Carter constant. Our conclusion is  that a physically relevant observer cannot detect violations of the Killing identity in the frozen star spacetime up to $\varepsilon^2$  corrections. Similarly, one can show that, to leading perturbative order,  the Carter constant is indeed constant.  The condition for the Carter constant being constant is $\;u^a\nabla_a C=0\;$, as this is what gives rise to the Killing identity~(\ref{kill}).

We have also verified that the Carter constant is  the separation constant in the Hamilton--Jacobi equation, $\;p^ap_a+ k=0\;$. This leads to the following pair of separated
equations,
\bea
\varepsilon^2 p_r^2 +\frac{r^2}{(a^2+r^2)^2}L^2-\frac{1}{\varepsilon^2}{\cal E}^2 +k &=& -{\widetilde C}\;, \\
\frac{1}{\Sigma}p^2_{\theta} + \frac{a^2 \cos^2{\theta}}{(a^2+r^2)^2}L^2&=& +{\widetilde C}\;,
\eea
where $p_a$ is a component of 4-momentum per unit mass and ${\widetilde C}$ is the Carter constant per square of the unit mass. The latter equation is telling us that ${\widetilde C}$ is typically a number of order unity,
which allows the former equation to be reexpressed  (using $\;p_a=g_{ab}u^b$) into
 Eq.~(\ref{almostnull}).

\section{A rotating frozen star: The matter}

\subsection{The Einstein tensor}

The Einstein tensor $\;G_{\mu\nu}={\cal R}_{\mu\nu}-\frac{1}{2} g_{\mu\nu}{\cal R} \;$ or, equivalently, the stress tensor for the interior metric~(\ref{diagonal2}) takes a relatively simple form to leading order in $\varepsilon^2$. The diagonal components
at this order go as (with $\chi$ now relabeled as $t$)
\be
G^{t}_{\;\;t}\;=\;G^{r}_{\;\;r}\;=\; -\frac{\left(r^2+a^2\right)\left(r^2-3a^2\cos^2{\theta}\right)}{\left(r^2+a^2\cos^2{\theta}\right)^3}\;,
\label{GttGrr}
\ee
\be
G^{\theta}_{\;\;\theta}\;\;,\; \;G^{\phi}_{\;\;\phi}\;=\; {\cal O}[\varepsilon^2]\;.
\ee
As expected, the first two diagonal components are equal, meaning that $\;p=-\rho\;$, while the angular components are perturbatively small. Although this latter pair depend differently on $r$ and $\theta$, they both scale as $a^2$ and
thus go to zero in the non-rotating limit, as they must to agree with the static case.
Both $G^{t}_{\;\;t}$ and  $\;G^{r}_{\;\;r}$ are mildly divergent in the region $r\to 0$. This is similar to the non-rotating case \cite{trajectory} and, as previously discussed, can readily  be dealt with by using an appropriate regularization procedure.

Meanwhile, there is a single  non-vanishing off-diagonal element,
\be
G_{r\theta}=\frac{2~ a^2r \sin{2\theta}}{\left(r^2+a^2\cos{\theta}\right)^2}\;,
\label{offd}
\ee
which correctly goes to 0 as $a^2$ does. If one chooses to diagonalize the $r$--$\theta$-sector, the result is
that the diagonal components are unchanged, up to $\varepsilon^2$ corrections, so that the off-diagonal component has minimal physical
impact.

The Einstein tensor,  although formally a four-dimensional tensor, is determined to leading order by only the two-dimensional angular part of the metric $\;g_{\theta\theta} d\theta^2 +g_{\phi\phi} d\phi^2\;$.
A related, interesting feature of the Einstein tensor is that it implies that $\;\rho=-\frac{1}{8\pi} G^{t}_{\;\;t}\;$  can be negative for small-enough values of $r$ because of the factor of $\;r^2-3a^2\cos^2{\theta}\;$ in Eq.~(\ref{GttGrr}). This apparent negative-energy region corresponds to parts of the interior where the extrinsic curvature on radial slices becomes negative. A  system, even a classical one,  with negative energy density is not necessarily a problem unto itself provided that the NEC remains satisfied
\cite{Zaslavskii}.   This is certainly true for our special class of infalling observers, for which the physically relevant measure of energy is, in fact, the radial null energy. Ironically enough, since the radial null energy vanishes throughout the interior,  these observers would believe that our non-vacuum source for the Kerr exterior is, in fact, a vacuum.

Although the RNEC is not in jeopardy,  what about the angular components of this condition? These are clearly violated in the negative-energy region since the transverse pressures (or angular components of $G^a_{\;\;\;b}$) go as $\;p_{\theta}\;,\; p_{\phi} \sim \varepsilon^2\;$ and, therefore,
cannot possibly compensate for a  negative value of $\rho$. Nevertheless, this is not a problem for the following reason: No  physically relevant  observer could ever observe a local violation of the  NEC to leading order in $\varepsilon^2$.  To elaborate, from Eq.~(\ref{motion}),  the squares of the (effective) angular velocities  are at most order $\varepsilon^2$ in comparison to the squares of the leading velocities ${\cal E}$ and $u^r$. For instance, allowing $(u^r)^2$ to be smaller than ${\cal E}^2$ by $\varepsilon^2$, compensating for this change with $\;(u^{\theta})^2 =\frac{\varepsilon^2}{\Sigma}\;$  and contracting the Einstein tensor twice with the modified velocity, one then obtains
\be
G^{a}_{\;\;b}u_au^b \;=\; -\left(G^t_{\;\;t} -  G^r_{\;\;r}\right)\;-\; \varepsilon^2  G^{r}_{\;\;r}  \;+\; 2{\cal E}\sqrt{\frac{\varepsilon^2}{\Sigma}}G_{r\theta}  \;+\;\varepsilon^4 G^{\theta}_{\;\;\theta}\;,
\ee
showing that, after diagonalization of the $r-\theta$ components of $G$,  any such violation is indeed $\varepsilon^2$ suppressed.

\subsection{Mass and angular momentum}

We will next  calculate both  the mass and the angular momentum of a  frozen star using the standard expressions \cite{HartleSharp,fourlaws,MTW}, which are also valid for an approximate, isolated horizon \cite{ashtekarbadri1,ashtekarbadri2,ashtekarbadri3,Hayward1,Hayward2}, such as the outer surface of a frozen star.  To fix  the ambiguity in the lapse $N$, we adopt   the areal radius $R$, as is done in the just-cited discussions on isolated horizons. Then the 3-volume $d^3V$ on the horizon  can be decomposed as $\;d^3 V = dR~ R^2 d\Omega\; $. We further simplify our expressions by working in units in which $\;R_+^2= r_+^2 + a^2= \dfrac{1}{4 \pi}A\;$, is set to  unity (but restore  the units at the end of the calculations).

In these units, the irreducible mass of a Kerr BH is  \cite{countfloyd,Christodoulou1,Christodoulou2,MTW}
\be
M_{irr}\;=\;\frac{1}{2} \sqrt{\frac{A}{4\pi}}= \frac{1}{2}\;,
\ee
which is the mass of a Schwarzschild BH with surface area $\;A= 4\pi R_+^2\;$ and
obviously in agreement  with the Schwarzschild mass in the $\;a\to 0\;$ limit.

Let us begin by using   the Einstein equation $\;-\frac{1}{8\pi}G^t_{\;\;t}=\rho\;$, to determine the mass of the frozen star.~\footnote{The mass calculation is from the perspective of an external observer, and so there is no contradiction with the notion that $\;\rho+p=0\;$  is the energy density from a local ZAMO's perspective.} The first step will be to calculate what we identify as the star's intrinsic rest mass $M_0$, which will turn out to be equal to $M_{irr}$,
\bea
8 \pi  M_0  &=&  -2\pi\int_0^{1} dR \int_0^{\pi}  d\theta  \sqrt{g_{\theta\theta}g_{\phi\phi}} \;G^t_{\;\;t}\;.
\label{add1}
\eea
 Note that one does not include $\sqrt{g_{rr}}$ in the measure as its positive contribution is canceled off  by the negative contribution of the gravitational potential energy~\cite{fourlaws,MTW,ashtekarbadri1,ashtekarbadri2,ashtekarbadri3,Hayward1,Hayward2}.

Performing the above integral,
\bea
   2 M_0   &=& \frac{1}{2}\int_0^{1} dR\; R^4 \int_0^{\pi} d\theta \sin{\theta}
\frac{\left(r^2-3a^2\cos^2{\theta}\right)}{\left(r^2+a^2\cos^2{\theta}\right)^3} \;=\; 1 \;,
\label{add2}
\eea
and then restoring the units, we find  that
\be
 M_0\;= \;\frac{1}{2} R_+ \;=\; M_{irr}\;.
\label{mirr}
\ee

The intrinsic angular momentum is similarly calculated via the standard techniques,
\be
8 \pi J_0 \;=\; -   2\pi\int_0^{1} dR \int_0^{\pi}  d\theta  \sqrt{g_{\theta\theta}g_{\phi\phi}}~ G^t_{~\phi}\;.
\label{angularM0}
\ee
However, a ZAMO finds that this angular momentum vanishes because $\;G^t_{~\phi}=0\;$, as expected.
So that, to calculate the angular momentum, we need to turn to a ``Boyer--Lindquist observer", for whom the frozen star is rigidly rotating at the horizon frequency $\Omega_H$, which in our current  units is $\;\Omega_H=a\;$. In which case, the metric contains a non-trivial  off-diagonal element,
\bea
ds^2&=&\left(\frac{a^2 R^4 \sin ^2\theta}{\Sigma}-\varepsilon ^2 \right)dt^2 \;-\;2a\sin ^2\theta \frac{ R^4 }{\Sigma} dtd\phi
\;+\; \frac{1}{\varepsilon ^2} dr^2\;+\; \Sigma d\theta^2\;+\;\frac{R^4 }{\Sigma}\sin ^2\theta d\phi^2\;. \nonumber \\
& &
\eea

The $t$--$t$ component of the Einstein tensor is unchanged,  while the desired $t$--$\phi$ component of the Einstein tensor is found to satisfy
\be
G^t_{~\phi}\;=\; a\; G^t_{~t}\;.
\ee
We can then use Eqs.~(\ref{add1}),~(\ref{angularM0}) and then~(\ref{add2}) to infer that
\be
 J_0 \;=\; a\; M_0 \;=\; \frac{1}{2} a\;.
\label{angularM1}
\ee
Or, with units restored,
\be
J_0\;=\; a\; M_{irr}\;.
\ee
The conclusion is that $\;a =J_0/M_0\;$ is indeed  the spin parameter of the frozen star.

Finally, following the usual procedure, one is instructed to include the rotational energy $E_{rot}$, to obtain the total gravitational mass $M$, of the rotating frozen star. The standard prescription for this is to integrate the relation $\;d E_{rot} = \Omega dJ\;$
and then sum the squares of the two energy contributions
\cite{Christodoulou1,MTW}, which leads
to the result
\bea
 M^2 &=&  \; M_{irr}^2 + \frac{1}{4}\frac{J^2}{M_{irr}^2}\;.
 \eea
Since the irreducible mass and the spin parameter of the frozen star are the same as  the corresponding  Kerr BH, its total mass is also the same.
The total angular momentum of the BH follows directly, albeit trivially, as
$\;J\;=\; a M \;$.

And so we find that the total gravitational mass and angular momentum of our rotating frozen star are precisely the same as those of a Kerr BH with
the same size and spin parameter. Our conclusion is that the interior matter in the frozen star is a possible source for the geometry of the  Kerr exterior metric.

\section{Conclusion}

We have identified a possible source for the Kerr geometry by incorporating rotation into our model for a regular BH mimicker --- the frozen star. Key properties of our solution include the regularity  of the metric and the Einstein tensor throughout the interior, except for a mild removable singularity at the center of the star, as well as the saturation of the RNEC, again,  throughout the interior. Our calculations were done to leading order in $\varepsilon^2$; however, we expect that they can be extended to higher order if necessary.

The above properties and the fact that the rotating star limits to the static star as the rotation is switched off are all important consistency checks. The intrinsic mass of our solution was calculated and found to be equal to the irreducible mass of the corresponding Kerr BH. We also calculated the angular momentum of the solution and found that the ratio of  the angular momentum to the mass is the spin parameter $a$, just as
it is for the corresponding Kerr BH. Together these findings mean that the total gravitational mass {\em and} the total angular momentum of our solution are equal to those of a  Kerr BH of the same size and with the same spin parameter for any acceptable value of $a$.

Having a rotating frozen star solution at hand is a key step toward our longer-range goal of calculating the gravitational-wave spectrum for the fluid modes of a deformed frozen star; for the analogous  static treatment, see \cite{fluctuations}. Rotation is  of course essential if one wants to use the current and soon-to-be available  observational data to seriously test the validity of any alternative BH model.  The lesson from \cite{fluctuations} is that the key to generating gravitational waves in an otherwise  ultrastable compact object is to deform the star in a way that modifies the saturating condition $\;G^t_{\;\;t}=G^r_{\;\;r}\;$.  We expect to report on the spectra for the rotating model in the near future, and then find the corresponding Love numbers and spectrum of ringdown modes.

\section*{Acknowledgments}
We thank Badri Krishnan and Yoav Zigdon for discussions. The research is supported by the German Research Foundation through a German-Israeli Project Cooperation (DIP) grant ``Holography and the Swampland'' and by VATAT (Israel planning and budgeting committee) grant for supporting theoretical high energy physics.
The research of AJMM received support from an NRF Evaluation and Rating
Grant 119411 and a Rhodes  Discretionary Grant SD07/2022. AJMM thanks Ben Gurion University for their hospitality during his visit.

\appendix

\section{The ADM form of Boyer-Lindquist}

Here, we want to verify that the ADM form of the time--time component in the metric~(\ref{metricX}) does indeed agree with
 its Boyer--Lindquist counterpart at least for the near-horizon solution, which is the relevant region in our analysis. Let us then recall the Boyer--Lindquist metric,
\bea
\left.ds^2\right|_{BL}&=& -\frac{\Delta-a^2\sin^2{\theta}}{\Sigma}dt^2 \;+\;\frac{\Delta}{\Sigma}dr^2
 \;-\;\frac{4 M r\, a  \sin^2{\theta}}{\Sigma} dt d\phi \nonumber \\
&+& \Sigma d\theta^2 \;+\; \left[r^2+a^2+\frac{2M r a^2 \sin^2{\theta}} {\Sigma}\right]\sin^2{\theta}d\phi^2\;,
\label{Kerr}
\eea
with the time--time component displayed  in a way which makes it clear that
the objective is to show that
\be
\omega^2 g_{\phi\phi}\;=\; \frac{g_{\phi t}^2}{g_{\phi \phi}}\; = \;\frac{a^2 \sin^2{\theta}}{\Sigma}\;.
\ee

Let us now substitute the relevant components of the Boyer--Lindquist metric into the middle term
in the above equation,
\be
 \frac{g_{\phi t}^2}{g_{\phi \phi}}\;=\; \frac{4M^2r^2 a^2 \sin^2{\theta}}{\Sigma\left[\left(r^2+a^2\right)\left(r^2+a^2\cos^2{\theta}\right) + 2Mra^2\sin^2{\theta}\right]}\;.
\ee

One can then use the relation $\;\Delta=r^2+a^2-2Mr\;$ along with $\;\sin^2{\theta}+\cos^2{\theta}=1\;$  to rewrite this
as
\be
 \frac{g_{\phi t}^2}{g_{\phi \phi}}\;=\; \frac{\left(r^2 +a^2-\Delta\right)^2   a^2 \sin^2{\theta}}{\Sigma\left[\left(r^2+a^2\right)^2-\Delta a^2 \sin^2{\theta}\right]}\;. \label{samesame}
\ee

Now one finds that taking  $\;\Delta \to 0\;$ on the right-hand side  leads to the desired outcome,
\be
 \frac{g_{\phi t}^2}{g_{\phi \phi}}\;=\; \frac{a^2\sin^2{\theta}}{\Sigma}\;.
\label{same}
\ee

To further see that the location of  the ergosphere also comes out correctly, one  can similarly substitute
 $\;\Delta=a^2\sin^2{\theta}\;$ on the right-hand side of Eq.~(\ref{samesame}) to again obtain Eq.~(\ref{same}).

\end{document}